# 180° Ferroelectric Stripe Nanodomains in BiFeO$_3$ Thin Films

Zuhuang Chen,[†,*] Jian Liu,[‡,§] Yajun Qi,[∥] Deyang Chen,[†] Shang-Lin Hsu,[†] Anoop R. Damodaran,[†] Xiaoqing He,[#] Alpha T. N'Diaye,[∇] Angus Rockett,[#,₶] and Lane W. Martin[†,§,₶*]

[†] Department of Materials Science and Engineering, University of California, Berkeley, Berkeley, CA 94720, USA

[‡] Department of Physics, University of California, Berkeley, California 94720, USA

[§] Materials Science Division, Lawrence Berkeley National Laboratory, Berkeley, CA 94720, USA

[∥] Hubei Collaborative Innovation Centre for Advanced Organic Chemical Materials, Key Laboratory of Green Preparation and Application for Materials, Ministry of Education, Department of Materials Science and Engineering, Hubei University, Wuhan 430062, P. R. China

[#] Department of Materials Science and Engineering and Materials Research Laboratory, University of Illinois, Urbana-Champaign, Urbana, IL 61801, USA

[∇] Advanced Light Source, Lawrence Berkeley National Laboratory, Berkeley, California 94720, USA

[₶] International Institute for Carbon Neutral Research, 744 Motooka, Nishi-ku, Fukuoka 819-0395, Japan

**Abstract**

There is growing evidence that domain walls in ferroics can possess emergent properties that are absent in bulk materials. For example, 180° ferroelectric domain walls in the ferroelectric-antiferromagnetic BiFeO$_3$ are particularly interesting because they have been predicted to possess a range of intriguing behaviors; including electronic conduction and enhanced magnetization. To date, however, ordered arrays of such domain structures have not been reported. Here, we report the observation of 180° stripe nanodomains in (110)-oriented $BiFeO_3$ thin films grown on orthorhombic $GdScO_3$ $(010)_O$ substrates, and their impact on exchange coupling to metallic ferromagnets. Nanoscale ferroelectric 180° stripe domains with $\{11\bar{2}\}$ domain walls were observed in films < 32 nm thick to compensate for large depolarization fields. With increasing film thickness, we observe a domain structure crossover from the depolarization field-driven 180° stripe nanodomains to 71° ferroelastic domains determined by the elastic energy. Interestingly, these 180° domain walls (which are typically cylindrical or meandering in nature due to a lack of strong anisotropy associated with the energy of such walls) are found to be highly-ordered. Additional studies of $Co_{0.9}Fe_{0.1}/BiFeO_3$ heterostructures reveal exchange bias and exchange enhancement in heterostructures based-on $BiFeO_3$ with 180° domain walls and an absence of exchange bias in heterostructures based-on $BiFeO_3$ with 71° domain walls; suggesting that the 180° domain walls could be the possible source for pinned uncompensated spins that give rise to exchange bias. This is further confirmed by X-ray circular magnetic dichroism studies, which demonstrate that films with predominantly 180° domain walls have larger magnetization than those with primarily 71° domain walls. Our results could be useful to extract the structure of domain walls and to explore domain wall functionalities in $BiFeO_3$.

Epitaxial ferroelectric thin films usually form domains in order to minimize the total free energy of the system.[1] More specifically, ferroelectric 180° domains form to compensate the depolarization field resulting from imperfect screening of the polarization charge[2, 3] and non-180° ferroelastic domains generally form to accommodate elastic energy due to elastic film-substrate interaction (*i.e.*, lattice-mismatch, epitaxial strain).[4,5] Thus the domain structure of a ferroelectric film is determined by minimizing the sum of the electrostatic depolarization energy, the elastic strain energy, and the domain wall energy, and the resulting domain structure has a profound impact on the dielectric permittivity, piezoelectric response, and polarization switching behavior.[1,6] For instance, both theoretical and experimental studies show that ferroelectric thin films with 180° nanodomains possess an enhanced dielectric response.[7-9] Furthermore, recent studies have shown that domain walls themselves can possess additional functionalities (*e.g.*, electronic conductivity) and have the potential for other interesting effects; making domain walls potential candidates for active device elements in future nanoelectronics.[10-13] Therefore, it is important to deterministically control the domain structure and to understand the structure and properties of domain walls in ferroelectric films.

Among ferroelectrics, $BiFeO_3$ has been extensively studied in recent years due to its room-temperature multiferroism, large polarization, high Curie temperature, relatively low band gap, and rich strain-temperature phase diagram; all of which make it an attractive candidate for applications.[14] In the bulk, $BiFeO_3$ has a rhombohedral structure and the polarization is oriented along $<111>$ (note that pseudocubic indices are used throughout, unless otherwise specified), leading to three types of possible domain walls separating regions with polarization orientations differing by 71°, 109°, and 180°. Charge neutrality and mechanical compatibility conditions impose constraints on the orientations of these domain walls: equilibrium uncharged 71° and 109° domain walls occur on {101} and {100}, respectively; while uncharged 180° domain walls form in crystallographic planes parallel to the polarization vectors inside the adjacent domains. Periodic 71° and 109° stripe domains have been obtained in $BiFeO_3$ epitaxial thin films,[15-17] and thus make $BiFeO_3$ a model system to study structure and properties of domain walls in ferroic systems.[11,18,19] Stripe arrays of 180° domain walls in $BiFeO_3$, however, have not been observed. This is particularly disappointing since the 180° domain walls have been predicted to exhibit the largest band gap reduction and magnetization out of the three possible domain wall variants in $BiFeO_3$.[12,20] Moreover, the structure and properties of 180° ferroelectric domain walls have been extensively studied theoretically in recent years;[20,21] and a non-Ising character of the 180° domain wall, arising from flexoelectric effects, has been proposed.[22] Therefore, experimental verification of the existence of periodic 180° stripe domains in $BiFeO_3$ is important for better understanding the domain formation mechanism, domain wall structure, and its contributions to magnetic and electric properties in this multiferroic material.

In this letter, we study the structure and properties (and their thickness-evolution) of periodic arrays of 180° domain walls in epitaxial (110)-oriented $BiFeO_3$ thin films grown on orthorhombic $GdScO_3$ $(010)_O$ single-crystal substrates (where the subscript "O" denotes an orthorhombic index). The 180° stripe nanodomains with $\{11\bar{2}\}$ domain walls form in ultrathin films to compensate large depolarization fields. With increasing film thickness, the films undergo a crossover from the 180° stripe domains to 71° ferroelastic domains. The 71° ferroelastic domains (like those observed in the growth of BiFeO$_3$ on SrTiO$_3$ (110) substrates) appear in thicker films where the elastic energy begins to dominate the domain formation. Additional studies of Co$_{0.9}$Fe$_{0.1}$/BiFeO$_3$ heterostructures reveal exchange bias and exchange enhancement in heterostructures based-on BiFeO$_3$ with 180° domain walls and negligibly small exchange bias in heterostructures based-on BiFeO$_3$ with predominantly 71° domain walls; suggesting that the 180° domain walls could be the possible source for pinned uncompensated spins that give rise to exchange bias. Subsequent X-ray magnetic circular dichroism (XMCD) studies reveal that BiFeO$_3$ films possessing 180° domain walls have larger dichroism (and thus magnetization) than those possessing 71° domain walls.

BiFeO$_3$ films with thicknesses ranging from 10-80 nm were deposited by pulsed-laser deposition following established procedures[16] on orthorhombic GdScO$_3$ (010)$_O$ single-crystal substrates (with lattice constants $a_0$ = 5.488 Å, $b_0$ = 5.746 Å, and $c_0$ = 7.934 Å).[23] The orthorhombic unit cell can be described with a pseudocubic space group, in which the $[001]_O$, $[100]_O$, and $[010]_O$ orthorhombic directions correspond to the $[001]$, $[1\bar{1}0]$, and $[110]$ pseudocubic directions, respectively. Therefore, the GdScO$_3$ (010)$_O$ substrates are akin to a (110)-oriented cubic perovskite substrate and are likely to result in epitaxial growth of (110)-oriented films. It should be noted, however, that this results in a large anisotropy in the values for the lattice mismatch along the two in-plane directions: -2% (compressive) and 0.05% (tensile) along $[100]_O$ and $[001]_O$, respectively. Detailed structural information was obtained using high-resolution X-ray diffraction (XRD, XPert MRD Pro, Panalytical). A representative $\theta - 2\theta$ XRD pattern of a ~32 nm thick BiFeO$_3$ / GdScO$_3$ (010)$_O$ heterostructure [Fig. 1(a)] reveals only *hh*0 diffraction peaks for the film and the substrate suggesting epitaxial growth without impurity phases. The thickness fringes apparent near the BiFeO$_3$ diffraction peaks indicate the high quality of the films. The out-of-plane lattice parameter of the film is measured to be ~2.83 Å, larger than the (110) *d*-spacing of bulk BiFeO$_3$ ($d_{110,bulk}$ = 2.804 Å), suggesting that the film is under average in-plane compressive strain.

Surface morphology and piezoresponse force microscopy (PFM) investigations were completed using a scanning probe microscope (Cypher, Asylum Research). A typical atomic force microscopy topographic image of the same film [Fig. 1(b)] further reveals a smooth surface with a root-mean-square (RMS) roughness of only ~500 pm (across a 3 x 3 μm area). In rhombohedral BiFeO$_3$, there are eight possible energetically-degenerate ferroelectric polarization variants. The average in-plane compressive strain imposed by the GdScO$_3$ substrate, however, makes the four domain variants with polarization lying within the (110) (*i.e.*, the growth plane) energetically unfavorable and thus reduces the number of possible polarization variants to four (two pointing into the plane of the film and two pointing out of the plane of the film) thereby allowing for simplified domain analysis. Characteristic out-of-plane (*i.e.*, vertical phase) [Fig. 1(c), inset shows corresponding amplitude image] and in-plane (*i.e.*, lateral phase) [Fig. 1(d), inset shows corresponding amplitude image] PFM images, taken with the cantilever's long and scan axes along $[100]_O$ (*i.e.*, $[1\bar{1}0]$) of the ~32 nm thick film, are provided. Stripe-like contrasts with domain walls lying along the $[100]_O$ are clearly observed. The alternating bright and dark contrast in the vertical phase images signifies a corresponding change in the out-of-plane component of the polarization vector. The lateral phase image also reveals a 180° phase difference, with uniform amplitude of the adjacent domains, which suggests that the in-plane polarization component in the two domains are pointing to the left (*i.e.*, [001]) and right (*i.e.*, [00$\bar{1}$]) with respect to the cantilever's long axis. Based on the lateral and vertical PFM studies, the observed domain structure is attributed to a periodic, stripe array of domains with a 180° difference in their polarization directions separated by 180° domain walls that that lie on {11$\bar{2}$} and intersect with the film surface along $[100]_O$. Such domain structures are indicative of the domain structures observed at multiple locations on all samples (with film thickness < ~35 nm) probed; at no time have we observed the domain wall structures to meander or diverge from the parallel structures observed here.

To confirm these observations and the nature of the domain structure, cross-sectional transmission electron microscopy (TEM) studies were completed. TEM specimen were prepared using standard procedures consisting of cutting, gluing, mechanical polishing, and ion milling, and were studied on a Tecnai G$^2$ 20 microscope at 200 kV. At first glance, one might expect to see no contrast from 180° ferroelectric domains because there is no difference in the unit cell on either side of the domain wall. It has been shown, however, that contrast can be obtained in dark-field TEM imaging due to the failure of Friedel's law (*i.e.*, as a result of a dynamical diffraction effect).[24] Dark-field cross-sectional TEM images of the ~32 nm thick film taken along the $[100]_O$ zone axis under two beam conditions with, g = $[010]_O$ (or [110]) [Fig. 2(a)] and g = $[0\bar{1}0]_O$ (or [$\bar{1}\bar{1}$0]) [Fig. 2(b)], reveal alternating dark and bright contrasts indicative of the expected domain structure. The contrast reversal in the two images demonstrates that they are indeed ferroelectric domains. As expected, the domain walls intersect with the substrate-film interface at an angle of ~55°, which is close to the angle between the domain wall plane [*i.e.*, $(11\bar{2})$] and plane of the film [*i.e.*, (110)] – confirming the plane expected for 180° walls. A selected area electron diffraction (SAED) pattern taken from the film region including the domain walls [Fig. 2(c)], shows no splitting even on high-order diffraction spots (Supporting Information, Fig. S1), indicating that there are no ferroelastic domains in the region. Thus, combining the PFM and TEM observations, we conclude that the domain structure of the $BiFeO_3$ films (<~ 35 nm thick) grown on $GdScO_3$ $(010)_O$ substrates are made up of arrays of 180° stripe domains with uncharged domain walls lying on the $(11\bar{2})$ [shown schematically, Fig. 2(d)].

Beyond the PFM and TEM studies, X-ray scattering was used to probe the periodic domain structures since regular domains can produce satellite peaks about the Bragg reflections of the film at specific reciprocal lattice vectors which have a component parallel to the relative polar shift directions.[3, 25-28] These satellite peaks provide rich information on the domain wall orientation and polar symmetry of the ferroelectric film. Armed with this knowledge and to circumvent the resolution limit of PFM, we have gone on to use detailed XRD reciprocal space mapping (RSM) studies to further probe the structure of the films. Both specular and off-specular RSMs are provided for a ~14 nm thick film, which is characteristic of the observations of all films with thickness <32 nm [Fig. 3] (for additional information see the Supporting Information, Fig. S2). Note also that the average domain width does scale with the film thickness, but the overall features are consistent. The first point to note is that from all RSM studies, the Bragg peaks from the film and substrate possess the same diffraction peak position in the in-plane directions (*i.e.*, $Q_x$- and $Q_y$-values) which confirms that the thin film is coherently strained to the substrate despite the rather large in-plane structural anisotropy. Upon closer inspection, however, only two Bragg spots from the substrate and the film are detected in the $(HK0)_O$ when the incident X-ray beam is along $[100]_O$ [Figs. 3(a) and (b)]. When the incident X-ray is aligned parallel to the $[001]_O$, up to second-order satellite peaks are observed in maps about both the $020_O$- and $042_O$-diffraction conditions of the GdScO$_3$ substrate [Figs. 3(c) and (d), respectively] indicating that the nanodomains are well ordered along $[100]_O$ (*i.e.*, $[1\bar{1}0]$) with a narrow domain size distribution. The spacing between the first-order satellite and central Bragg peaks yields a domain periodicity of ~20 nm (again, this is for the 14 nm thick film, which is close to or below the standard resolution for PFM analysis and less than that observed for the 32 nm thick film and consistent with Kittel's law for domain scaling).[29,30] The presence of satellite peaks in the $020_O$-diffraction condition (*i.e.*, the 110-diffraction condition) indicates that the out-of-plane polarization is periodically modulated (*i.e.*, the out-of-plane component of the polarization in the alternating domains point in opposite directions). This eliminates the possibility of a structure consisting of 71° ferroelastic domains which have the same out-of-plane polarization component, as typically observed in (110)-oriented BiFeO$_3$ films grown on SrTiO$_3$ substrates.[31] Therefore, the XRD RSM studies further confirm that the domain structure in the ultrathin films is that of the 180° stripe domains with domain walls lying along in-plane $[001]_O$ (consistent with the observations from both PFM and TEM as obtained for slightly thicker films).

Such 180° domain walls are ferroelectric in nature (*i.e.*, possess no elastic energy associated with them) and the equilibrium orientation of such domain walls is subject to charge compatibility and domain wall energy. Additionally, the requirement of charge neutrality dictates that the wall should be parallel to the spontaneous polarization vectors on either side of that domain wall. For instance, uncharged 180° domain walls in tetragonal and rhombohedral ferroelectrics can exist on any crystallographic plane in the [001] and [111] zones, respectively. Deviation from the ideal orientation of the domain wall may induce charging in the domain walls, which is energetically costly.[1] If the anisotropy of the 180° domain walls energy is small (*i.e.*, the dependence of the domain wall energy on orientation is small), as is the case (to some extent) in the prototypical ferroelectrics $PbTiO_3$[3,8,9,27,28] and $BaTiO_3$,[30] these domains are cylindrical in shape and the domain boundaries are not confined to a given crystallographic direction or plane thus giving rise to 180° domain walls that can meander in a random fashion.[32, 33] In this work, however, the 180° domain walls in the $BiFeO_3$ / $GdScO_3$ $(010)_O$ heterostructures are highly-ordered and straight (*i.e.*, the domains are slab shaped). This indicates that there is a large anisotropy associated with the different possible variants of 180° domain walls in $BiFeO_3$. Crystallographically prominent charge-neutral walls in rhombohedral ferroelectrics are either ($1\bar{1}0$)- or ($11\bar{2}$)-type. Our experimental results reveal the preference of the system to form $\{11\bar{2}\}$ domain walls. To the best of our knowledge, this is the first experimental observation of periodic 180° domains with $\{11\bar{2}\}$ domain walls in rhombohedral ferroelectrics. We note that most previous theoretical studies on the structure and properties of 180° domain walls in rhombohedral ferroelectrics focus on $\{1\bar{1}0\}$-type 180° walls.[20,21] The ultimate driving force for the selection of that the $\{11\bar{2}\}$ domain wall plane could be influenced by a number of factors in $BiFeO_3$ including the added energy arising from antiferromagnetic ordering and antiferrodistortive distortions in $BiFeO_3$[20,34] and/or epitaxial constraint from the substrate which could alter the preferred domain wall orientation even for 180° domain walls.[35] We note that recent theoretical studies proposed that, besides mechanical compatibility and electrical neutrality conditions, a rotational compatibility condition could also play an important role determining the orientation of permissible domain walls in perovskites with oxygen octahedral tilt instability.[34]

Having established the presence and nature of 180° domain walls in the $BiFeO_3$ films, we proceeded to probe the thickness-dependent evolution of these features. Analysis of both the vertical [out-of-plane, Fig. 4(a)] and lateral [in-plane, Fig. 4(b)] PFM phase images of a 70 nm thick film reveals that by this thickness, the vast majority of the sample possesses uniform vertical phase contrast (suggesting that the film is primarily poled downward) while the lateral phase reveals the presence of large, irregular in-plane domains. These two items taken together indicate that the domain structure is dominated by two ferroelastic domains separated by irregular 71° domain walls; similar to that typically observed in $BiFeO_3$/$SrTiO_3$ (110) heterostructures.[31] The formation of 71° ferroelastic domains in the 70 nm thick films is further confirmed by RSM studies taken about the $042_O$- (*i.e.*, 221) and $240_O$- (*i.e.*, 310) diffraction conditions of the $GdScO_3$ substrate [Fig. 4(c) and (d), respectively]. The corresponding 221- and 310-diffraction conditions of the $BiFeO_3$ split into two and one peak, respectively, again consistent with what is observed for films grown on $SrTiO_3$ (110) substrates with two polarization variants separated by 71° domain walls.[31,36] In addition, the position of the 221-diffraction condition of the $BiFeO_3$ along the $[001]_O$ (*i.e.*, [001]) has identical in-plane position with that of the corresponding in-plane diffraction condition of the $GdScO_3$, while the 310-diffraction condition of the $BiFeO_3$ peak has a smaller $Q_x$ value than that of the corresponding diffraction condition of the $GdScO_3$ substrate along $[100]_O$ (*i.e.*, [1$\bar{1}$0]). That is, the BiFeO$_3$ films are uniaxially strained (*i.e.*, coherently strained along the [001]$_O$ (*i.e.*, [001]), but relaxed (or partially strained) along the [100]$_O$ (*i.e.*, [1$\bar{1}$0]) as is typically observed in (110)-oriented perovskite films).[35] From the RSMs, we could extract the lattice parameters for the BiFeO$_3$ film to be: $a_m$ = 3.967 Å, $b_m$ = 5.568 Å, $c_m$ = 5.632 Å, and $\theta$ = 89.2°.

Ultimately, the formation and stability of the domain configuration of a ferroelectric film depends on the electrostatic and elastic boundary conditions (*i.e.*, depolarization field and elastic strain). The depolarization field (for films with uniform out-of-plane polarization) is inversely proportional to the film thickness; and could be reduced by free charges from metallic electrodes,[38] ionized adsorbates,[39] charged point defects within the ferroelectric itself,[40] suppression of polarization,[41] polarization rotation toward in-plane,[42] or by formation of domains with alternating out-of-plane polarization sign.[3, 26,43] As shown above, in the ultrathin BiFeO$_3$ films on insulating GdScO$_3$ substrates, charged defects/carriers in the film or ionized adsorbates on the surface are seemingly insufficient to fully screen the large depolarization field and thus 180° domains form to decrease the depolarization energy. With increasing film thickness, the effect of depolarization field becomes weaker and elastic strain energy increases and thus ferroelastic domains form to relax the strain energy. Such a domain structure transition from the depolarization field energy-driven 180° domains to the elastic energy-driven 71° domains is further supported by a coexistence of 180° domains and 71° domains in films with an intermediate film thicknesses of 40 nm (Supporting Information, Figs. S3 and S4). Similar domain structure evolution with film thickness have been observed in tensile-strained, (001)-oriented PbTiO$_3$ thin films grown on DyScO$_3$ (110)$_O$ substrates, where 180° domains have been observed in ultrathin films (to compensate the large depolarization field)[26] and periodic *c/a* domain structures have been observed in relatively thicker films to compensate the increasing elastic energy.[44,45]

As noted above, such domain structures can play an important role in the evolution of material properties. The role of domain walls in impacting the evolution of exchange coupling with metallic ferromagnets has been studied in (001)-oriented $BiFeO_3$ thin films with mosaic domains.[46-48] With this in mind, we have probed the exchange coupling between $Co_{0.9}Fe_{0.1}$ layers and (110)-oriented $BiFeO_3$ films controlled to possess either 180° or 71° domain wall structures. For the exchange bias studies, heterostructures of 2.5 nm Pt / 2.5 nm $Co_{0.9}Fe_{0.1}$ were deposited *ex situ* in a 20 mT applied field on the $BiFeO_3$ films by DC sputtering with a base pressure of ~ $10^{-8}$ Torr at room temperature. Subsequently, magnetic properties were measured using a Quantum Design SQUID magnetometer. For heterostructures based on 30 nm thick $BiFeO_3$ films exhibiting ordered arrays of 180° domain walls, we observed both an exchange enhancement of the coercive field (~ 60 Oe, black and red curves, Fig. 5a) as compared to that of the $Co_{0.9}Fe_{0.1}/GdScO_3$ $(010)_O$ heterostructures (~ 15 Oe, blue curve, Fig. 5a) and an exchange bias (manifested as a shift of the hysteresis loop by ~ -25 Oe). In contrast, for heterostructures based on ~70 nm $BiFeO_3$ films with primarily 71° domain walls, only exchange enhancement of the coercive field (~ 50 Oe, Fig. 5b) is observed while there is negligible small exchange bias (< 3 Oe). These exchange bias studies suggest some intriguing differences in the magnetic behavior of the two types of domain structures. Exchange enhancement, common to both types of $BiFeO_3$ films, likely arises from a coupling of the ferromagnetic moments of the $Co_{0.9}Fe_{0.1}$ to the small canted moments due to the Dzyaloshinski-Moriya interaction (~ 0.02 $\mu_B$/Fe) in the bulk of the BiFeO$_3$ domain surface.[47] Exchange bias, on the other hand, requires the presence of pinned, uncompensated spins in the antiferromagnet.[46-48,49] The observed exchange bias in the heterostructures created on BiFeO$_3$ films with ordered arrays of 180° domain walls suggests the potential for such pinned, uncompensated spins at the domain walls.

Such observations are further supported by XMCD measurements carried out at beamline 6.3.1 of the Advanced Light Source, Lawrence Berkeley National Laboratory. These XMCD studies, focused on the Fe *L*-edge, used fixed circularly polarized X-rays and point-by-point reversal of the external magnetic field of magnitude 1.9 T at 30 K, in total electron yield configuration with grazing angle of 30° and incident beam along [100]$_O$ to probe the average magnetization of the sample surface. To ensure that the XMCD signal is of magnetic origin, we repeated the measurement with opposite polarization and confirmed that the asymmetry reverses. The X-ray absorption spectra (XAS) and corresponding XMCD at the Fe *L* edge are provided (Fig. 5c and d). The BiFeO$_3$ films exhibiting ordered arrays of 180° domain walls produce a normalized asymmetry of ~0.3% (Fig. 5c), while films exhibiting only 71° domain walls reveal no measurable asymmetry (Fig. 5d). From the ~ 0.3% XMCD asymmetric signal, and using the data from other iron oxide systems (*i.e.*, Fe$_3$O$_4$) and previous results on mixed-phase BiFeO$_3$ films;[51] we estimate a magnetic moment in the range of 10-15 emu/cc. From PFM and XRD results (Supporting Information, Fig. S2), the average 180° domain size of a ~ 30 nm thick film is around 15-20 nm. Taking the domain wall thickness of ~2 nm thick in BiFeO$_3$,[11] the volume fraction of the 180° walls is in the range of 10-12 %. Using this volume fraction and the existing canted moment from the BiFeO$_3$ domain surface of ~5 emu/cc (note that it is difficult to detect the canted moment of the BiFeO$_3$ by the XMCD technique due to the small magnitude of the moment),[48,49] the magnetic moment at the 180° walls is estimated to be potentially as large as 40-80 emu/cc. These results are consistent with prior reports which obtained quantitative estimates of the magnetization profile across a 180° domain wall in BiFeO$_3$ and suggested the possibility of enhanced moments at the domain walls of between 20-130 emu/cc.[12] It should also be noted that both exchange bias and XMCD studies of high-density, periodic arrays of 71° domain walls in (001)-oreinted BiFeO$_3$ films grown on DyScO$_3$ (110)$_O$ substrates also reveal negligible exchange bias and XMCD signals despite a much higher density of the domain wall features.[47,53] In turn, it appears that the 180° ferroelectric domain walls are different from 71° domain walls and are a possible source of the pinned, uncompensated spins which give rise to exchange bias and the enhanced XMCD signal.

In conclusion, 180° stripe nanodomains with $(11\bar{2})$ domain walls form in ultrathin films of BiFeO$_3$ to reduce the depolarization field when films are grown on insulating orthorhombic GdScO$_3$ (010)$_O$ substrates. With increasing film thickness, a crossover from the depolarization-field-driven 180° domains to strain-energy-driven 71° domains is observed. Our results demonstrate the key role of the electrostatic and elastic boundary conditions on the evolution of domain structure in ferroelectric films. Additionally, we further reveal that 180° domain walls are a possible source of the pinned, uncompensated spins responsible for exchange bias observed in Co$_{0.9}$Fe$_{0.1}$/BiFeO$_3$ heterostructures and enhanced magnetization in BiFeO$_3$ films. The observed near-equilibrium 180° stripe nanodomain patterns, in turn, enable us to study the structure of 180° ferroelectric walls and the wall contributions to functional properties (e.g., electronic, magnetic, and optical properties) in multiferroic BiFeO$_3$.

**Supporting Information Available**

The supporting information section provides information of SAED pattern analysis of the 32 nm film, XRD/RSM data of 20 nm and 32 nm films, and AFM/PFM and XRD/RSM data for a 40 nm thick film. This material is available free of charge via the Internet at http://pubs.acs.org.

AUTHOR INFORMATION

**Corresponding Author:**

Email: zuhuang@berkeley.edu

Email: lwmartin@berkeley.edu**Acknowledgements**

Z.H.C. and L.W.M. acknowledges the support of the Army Research Office under grant W911NF-14-1-0104 and the Air Force Office of Scientific Research under grant MURI FA9550-12-1-0471. J.L. acknowledges support from the Quantum Materials FWP, Office of Basic Energy Sciences, Materials Sciences and Engineering Division, of the U.S. Department of Energy under Contract No. DE-AC02-05CH11231. Y.Q. acknowledges support of the National Science Foundation of China under grant 11204069. A.R.D. acknowledges the support of the National Science Foundation under grants DMR-1149062 and DMR-1124696, respectively. Part of the16 work was completed at the Advanced Light Source which is supported by the Director, Office of Science, Office of Basic Energy Sciences, of the U.S. Department of Energy under Contract No. DE-AC02-05CH11231. A.R. and L.W.M. acknowledge support by the International Institute for Carbon-Neutral Energy Research (WPI I$^2$CNER), sponsored by the Japanese Ministry of Education, Culture, Sport, Science and Technology. We are grateful to Ruijuan Xu for the domain schematic drawing.

**Figures and Captions**

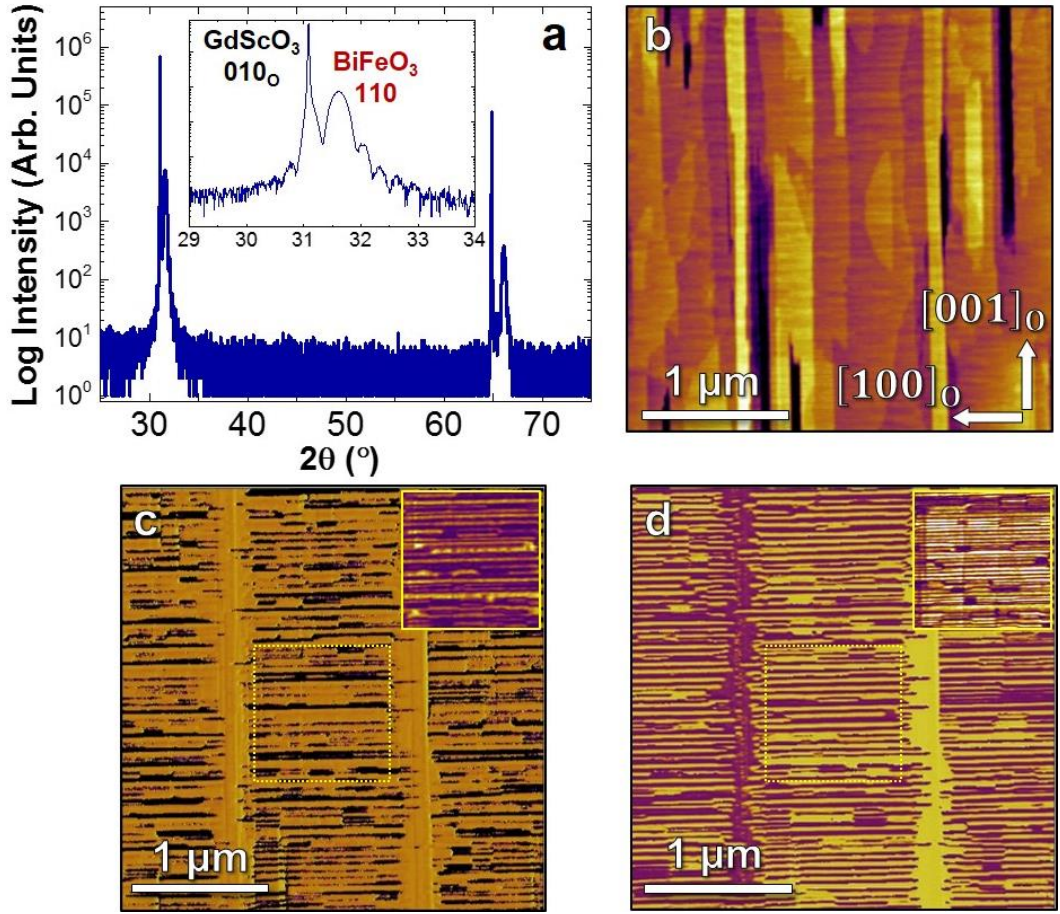

**Figure 1**. (a) XRD $\theta - 2\theta$ scan for a 32 nm BiFeO$_3$/GdScO (010)$_O$ heterostructure. Corresponding (b) atomic force microscopy topography image, (c) vertical (out-of-plane) phase (inset amplitude in yellow box) and (d) lateral (in-plane) phase (inset amplitude in yellow box) piezoresponse force microscopy images showing the domain structure of the same heterostructure.

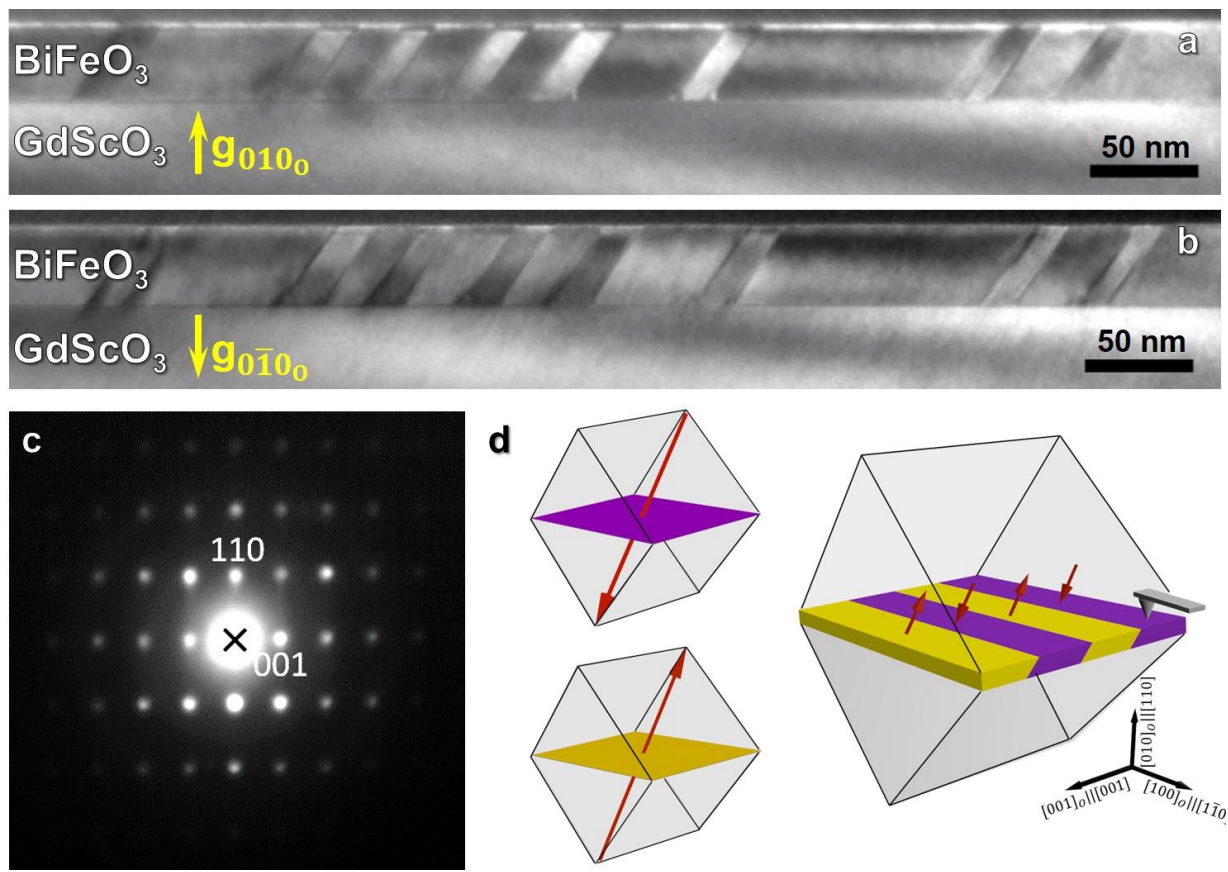

**Figure 2**. Dark-field transmission electron microscopy images of a 32 nm thick $BiFeO_3/GdScO_3$ $(010)_O$ heterostructure taken along the $[100]_O$ zone axis under two beam conditions (a) g = $[010]_O$ and (b) g = $[0\bar{1}0]_O$. (c) Selected area electron diffraction pattern taken along the $[100]_O$ zone axis from an area of the film including multiple domains. (d) Schematic illustration of the 180° domain structure, where the arrows represent the directions of the spontaneous polarization.

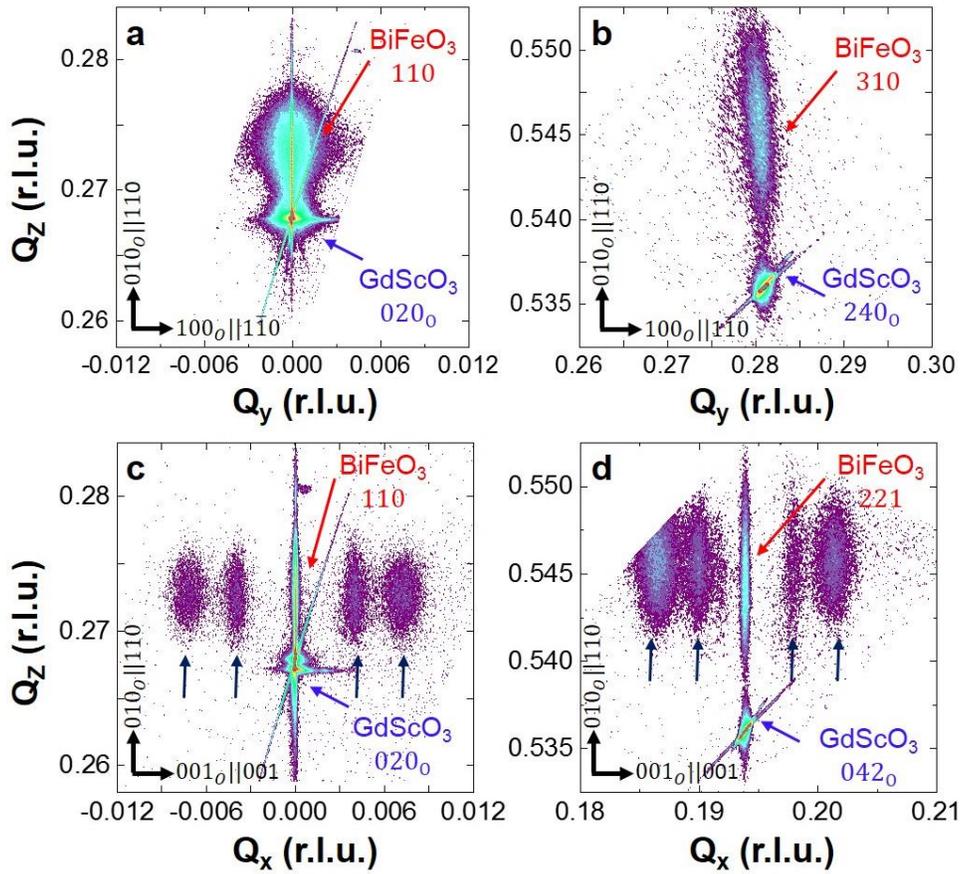

**Figure 3**. Reciprocal space mapping studies of a 14 nm thick BiFeO$_3$/GdScO$_3$ (010)$_O$ heterostructure about the (a) 020$_O$- (110-) and (b) 240$_O$- (310-) diffraction conditions of GdScO$_3$ (pseudocubic BiFeO$_3$) with the X-ray beam along the [100]$_O$ and studies about the (c) 020$_O$- (110-) and (d) 042$_O$- (221-) diffraction conditions of GdScO$_3$ (pseuodocubic BiFeO$_3$) with the X-ray beam incident along the [001]$_O$. The arrows indicate the positions of satellite peaks due to the periodic domains.

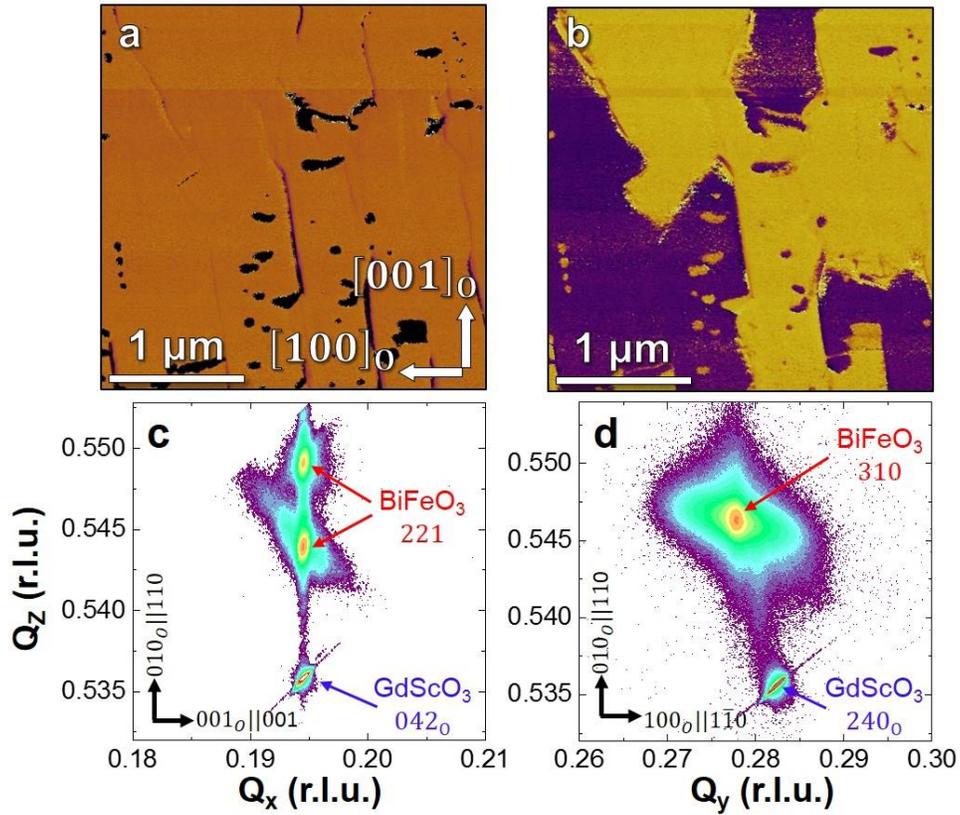

**Figure 4**. (a) Vertical (out-of-plane) and (b) lateral (in-plane) piezoresponse force microscopy phase images of a 70 nm thick $BiFeO_3/GdScO_3$ $(010)_O$ heterostructure. Corresponding reciprocal space mapping studies of the same film about the (c) $042_O$- ($\overline{2}21$-) and (d) $240_O$- ($310$-) diffraction conditions of $GdScO_3$ (pseuodocubic $BiFeO_3$).

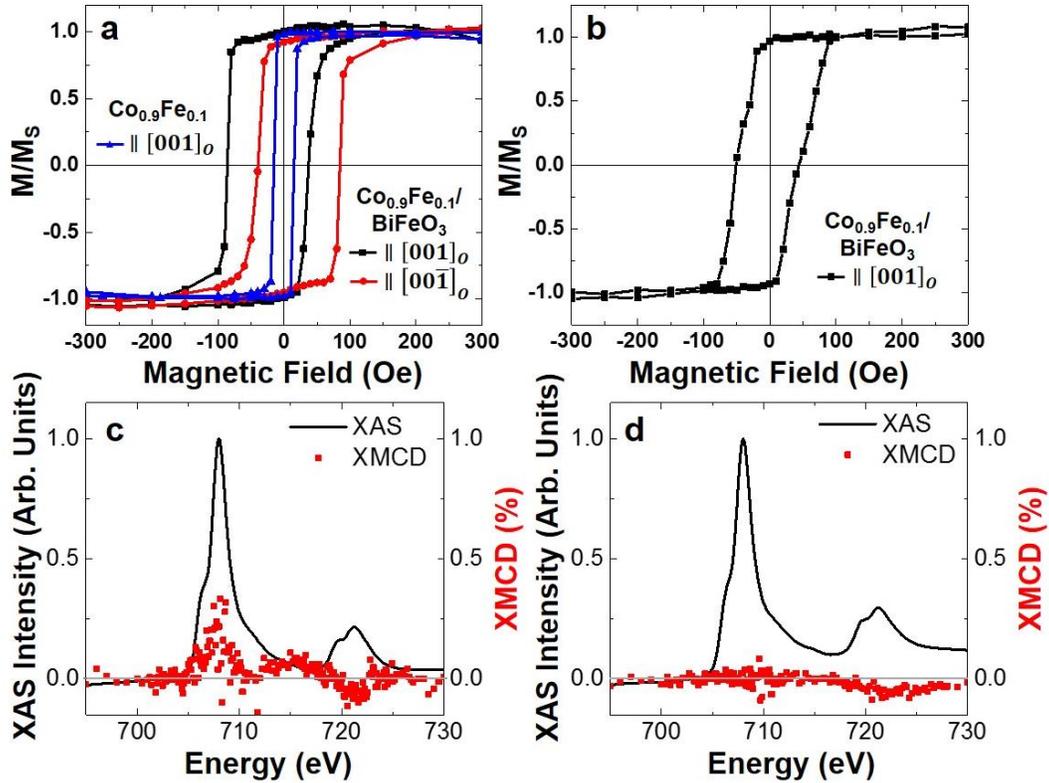

**Figure 5**. Room temperature magnetic hysteresis loops measured in-the-plane of the film for the Pt/$Co_{0.9}Fe_{0.1}$/$BiFeO_3$/$GdScO_3$ $(010)_O$ heterostructures (measured along in-plane $[001]_O$) where the $BiFeO_3$ films are controlled to possess (a) 180° stripe domains or (b) 71° ferroelastic domains. A loop of similarly grown Pt/$Co_{0.9}Fe_{0.1}$/$GdScO_3$ $(010)_O$ heterostructure is shown for comparison in (a). X-ray absorption spectroscopy (XAS) and X-ray magnetic circular dichroism (XMCD) spectra of the Fe $L_{2,3}$ edge taken for $BiFeO_3$ films controlled to possess (c) 180° stripe domains and (d) 71° ferroelastic domains reveals much stronger dichroism in films with 180° domain walls.